\documentclass{aims} 

\usepackage{amsmath}
\usepackage{amssymb}
\usepackage{bm}
\usepackage{booktabs}
\usepackage{paralist}
\usepackage[misc]{ifsym}
\usepackage{epsfig}
\usepackage{epstopdf}
\usepackage[colorlinks=true]{hyperref}
\hypersetup{urlcolor=blue, citecolor=red}
\allowdisplaybreaks

\def\currentvolume{X}
 \def\currentissue{X}
  \def\currentyear{200X}
   \def\currentmonth{XX}
    \def\ppages{X-XX}
     \def\DOI{10.3934/xx.xxxxxxx}

\DeclareMathOperator{\Tr}{Tr}

\newtheorem{theorem}{Theorem}[section]
\newtheorem{corollary}[theorem]{Corollary}

\newtheorem{lemma}[theorem]{Lemma}

\theoremstyle{definition}
\newtheorem{definition}[theorem]{Definition}
\newtheorem{remark}[theorem]{Remark}

\newtheorem{example}[theorem]{Example}

\title[Complex codebooks from functions]
{A generic construction of complex codebooks from
functions over Abelian groups}

\author[Z. Heng, C. Xie, J. Hu and X. Jing]{}



\thanks{$^*$Corresponding author: Chenkun Xie}

\begin{document}
\maketitle

\centerline{\scshape
Ziling Heng$^{1}$,
Chenkun Xie$^{{\href{mailto:13095999956@163.com}{\textrm{\Letter}}}*1}$,
Jiantao Hu$^{1}$
and Xiaoyan Jing$^{1}$}

\medskip

{\footnotesize
 \centerline{$^1$School of Science, Chang'an University, Xi'an 710064, China}
  
}

\bigskip

 \centerline{(Communicated by Handling Editor)}


\begin{abstract}
Codebooks with small inner-product correlation play a central role in direct spread code division multiple access communications, space-time coding, compressed sensing, and related areas because they improve system performance. This paper presents a generic construction and a modified construction of complex codebooks from functions on finite Abelian groups. By choosing perfect nonlinear functions, we systematically obtain several families of optimal or asymptotically optimal codebooks with respect to the Welch or Levenshtein bounds. These constructions recover a number of known optimal codebooks and also yield new codebooks with previously unreported parameters.
\end{abstract}


\section{Introduction}

In code division multiple access (CDMA) systems, codebooks (also called signal sets) with small inner-product correlation are usually used to distinguish among the signals of different users. An $(N,K)$ codebook $\mathcal{C}$ is a set
\[
\mathcal{C}=\{\bm{c}_{0},\bm{c}_{1},\ldots,\bm{c}_{N-1}\},
\]
where each codeword $\bm{c}_{l}$ $(0\leq l\leq N-1)$ is a unit norm $1\times K$ complex vector over some alphabet. The alphabet size is the number of elements in the alphabet. The maximum cross-correlation amplitude of an $(N,K)$ codebook $\mathcal{C}$ is defined by
\begin{equation}\label{eq:max-correlation}
I_{\max}(\mathcal{C})
=
\max_{0\leq i<j\leq N-1}
\left|\bm{c}_{i}\bm{c}_{j}^{H}\right|,
\end{equation}
where $\bm{c}^{H}$ denotes the conjugate transpose of a complex vector $\bm{c}$. $I_{\max}(\mathcal{C})$ is a performance measure of a codebook $\mathcal{C}$ in practical applications: the smaller its value, the weaker the interference among the signals and the better the multiple-access performance of the system. One important problem is to minimize the codebook's maximal cross-correlation amplitude, which is fundamental in the design of high-performance codebooks and in improving communication systems. Minimizing $I_{\max}(\mathcal{C})$ among the codewords of a codebook $\mathcal{C}$ can approximately optimize various performance metrics such as outage probability, average signal-to-noise ratio and symbol error probability for multiple-antenna transmit beamforming from limited-rate feedback \cite{love2003,mukkavilli2003}. In the context of unitary space-time modulation, minimizing $I_{\max}(\mathcal{C})$ is equivalent to minimizing the block error probability \cite{hochwald2000}. Moreover, minimizing $I_{\max}(\mathcal{C})$ of finite frames leads to minimal reconstruction error in multiple description coding over erasure channels \cite{strohmer2003}.

For a given $K$, we would construct an $(N,K)$ codebook with $N$ being as large as possible and $I_{\max}(\mathcal{C})$ being as small as possible simultaneously. However, the following Welch and Levenshtein bounds demonstrate a trade-off among the parameters $N$, $K$ and $I_{\max}(\mathcal{C})$ of a codebook $\mathcal{C}$.

\begin{lemma}[Welch bound \cite{welch1974}]\label{lem:welch}
For any $(N,K)$ codebook $\mathcal{C}$ with $N\geq K$,
\begin{equation}\label{eq:welch}
I_{\max}(\mathcal{C})
\geq
\sqrt{\frac{N-K}{(N-1)K}}.
\end{equation}
In addition, the equality in \eqref{eq:welch} is achieved if and only if
\[
\left|\bm{c}_{i}\bm{c}_{j}^{H}\right|
=
\sqrt{\frac{N-K}{(N-1)K}}
\]
for all pairs $(i,j)$ with $i\neq j$.
\end{lemma}

If a codebook $\mathcal{C}$ achieves the Welch bound in \eqref{eq:welch}, which is denoted by $I_{W}$, we call it a maximum-Welch-bound-equality (MWBE) codebook \cite{xia2005}. MWBE codebooks are employed in many applications including CDMA communications \cite{massey1993}, space-time codes \cite{tarokh2002} and compressed sensing \cite{tsiligianni2014}. In \cite{tsiligianni2014}, an MWBE codebook was investigated as an equiangular tight frame. To our knowledge, only the following constructions of MWBE codebooks were reported in the literature:
\begin{enumerate}
  \item[(1)] In \cite{sarwate1999,xia2005}, optimal $(N,N)$ and $(N,N-1)$ codebooks with $N>1$ were generated from the (inverse) discrete Fourier transform matrix or ideal two-level autocorrelation sequences.
  \item[(2)] In \cite{conway1996,strohmer2003}, optimal $(N,K)$ codebooks from conference matrices were given when either $N=2K=2^{d+1}$ with $d$ a positive integer or $N=2K=p^{d}+1$ with $p$ a prime number and $d$ a positive integer.
  \item[(3)] In \cite{ding2006,ding2007,xia2005}, optimal $(N,K)$ codebooks were constructed with cyclic difference sets in the Abelian group $(\mathbb{Z}_{N},+)$ or the additive group of finite fields or Abelian groups in general.
  \item[(4)] In \cite{fickus2012}, optimal $(N,K)$ codebooks from $(2,k,v)$-Steiner systems were presented.
\end{enumerate}

According to \cite{strohmer2003}, the Welch bound on $I_{\max}(\mathcal{C})$ of a codebook $\mathcal{C}$ is not tight when $N>K(K+1)/2$ for real codebooks and $N>K^{2}$ for all codebooks. The following Levenshtein bound turns out to be tighter than the Welch bound.

\begin{lemma}[Levenshtein bound \cite{levenshtein1983}]\label{lem:levenshtein}
For any real-valued codebook $\mathcal{C}$ with $N>K(K+1)/2$,
\begin{equation}\label{eq:levenshtein-real}
I_{\max}(\mathcal{C})
\geq
\sqrt{\frac{3N-K^{2}-2K}{(N-K)(K+2)}}.
\end{equation}
For any complex-valued codebook $\mathcal{C}$ with $N>K^{2}$,
\begin{equation}\label{eq:levenshtein-complex}
I_{\max}(\mathcal{C})
\geq
\sqrt{\frac{2N-K^{2}-K}{(N-K)(K+1)}}.
\end{equation}
\end{lemma}

In general, it is very hard to construct codebooks achieving the Levenshtein bound, which is denoted by $I_{L}$ (the right-hand side of \eqref{eq:levenshtein-real} or \eqref{eq:levenshtein-complex}). There are only a few constructions of codebooks achieving the Levenshtein bound. These codebooks meeting the Levenshtein bound were constructed from Kerdock codes \cite{calderbank1997,xiang2015}, perfect nonlinear functions \cite{ding2007signal}, bent functions over finite fields \cite{zhou2014}, and bent functions over Galois rings \cite{heng2017optimal}. Codebooks achieving the Levenshtein bound are employed in quantum physics and the design of spreading sequences for CDMA and sets of mutually unbiased bases \cite{ding2007signal,wootters1989}.

Since it is very difficult to construct optimal codebooks, there have been a number of attempts to construct codebooks asymptotically (or nearly) achieving the Welch bound or the Levenshtein bound \cite{heng2017new,hu2014,luo2018,tan2016,zhang2012}, i.e., $I_{\max}(\mathcal{C})$ is slightly higher than the Welch bound $I_{W}$ or the Levenshtein bound $I_{L}$, but asymptotically achieves one of them:
\[
\lim_{K\to\infty}\frac{I_{\max}(\mathcal{C})}{I_{W}}=1
\qquad\text{or}\qquad
\lim_{K\to\infty}\frac{I_{\max}(\mathcal{C})}{I_{L}}=1.
\]
In \cite{sarwate1999}, Sarwate gave some nearly optimal codebooks from codes and signal sets. In \cite{ding2006,ding2008,li2015,zhang2012,zhou2011}, the authors constructed some nearly optimal codebooks based on almost or relative difference sets. In \cite{yu2012}, Yu presented some nearly optimal codebooks from binary sequences. Heng and Chen constructed several nearly optimal codebooks from certain hybrid character sums \cite{heng2023}. Besides, Hu and Wu proposed new constructions of nearly optimal codebooks from difference sets and the product of Abelian groups \cite{hu2014}. Recently, Heng, Wang and Li constructed several families of 
asymptotically optimal codebooks by hybrid character sums from vectorial dual-bent 
functions \cite{heng2026hybrid}.

In this paper, A generic construction and a modified construction of complex codebooks from functions over Abelian groups are presented. Sufficient conditions under which the resulting codebooks are optimal or asymptotically optimal are provided. With these constructions, several families of optimal or asymptotically optimal complex codebooks, including new ones and known ones, are produced.

This paper is organized as follows. In Section~2, characters of Abelian groups, the discrete Fourier transform and perfect nonlinear functions are reviewed. In Section~3, the constructions are presented, and several optimal or asymptotically optimal codebooks are constructed with perfect nonlinear functions and other specific functions. Section~4 concludes the paper and discusses possible directions for future work.

\section{Preliminaries}

This section reviews some basic facts and results concerning characters on Abelian groups, Fourier transforms, and perfect nonlinear functions.

\subsection{Characters of Abelian groups}

Let $q$ be a power of a prime $p$, and let $(A,+)$ be a finite Abelian group. Put $A^{*}:=A\setminus\{0\}$, where $0$ is the identity element of $A$. We first recall the definition of a character on $A$. Let $e$ be the maximum order of an element of $A$; it is called the exponent of $A$. A character of $A$ is a homomorphism $\chi$ from $A$ to a multiplicative group, that is,
\[
\chi(a+a')=\chi(a)\chi(a')
\]
for all $a,a'\in A$.
All characters of $A$ form a group $\widehat{A}$ under multiplication, where for all $\chi,\chi'\in\widehat{A}$ and $a\in A$,
\[
(\chi\chi')(a)=\chi(a)\chi'(a).
\]
The inverse of a character $\chi\in\widehat{A}$ is defined by
\[
\chi^{-1}(a)=\chi(-a)=\overline{\chi(a)},
\]
where $\overline{\chi(a)}$ is the complex conjugate of $\chi(a)$. It is well known that $\widehat{A}$ is isomorphic to $A$ \cite{hewitt1979}. Hence, the number of characters of $A$ equals $|A|$. For each $\alpha\in A$, let $\chi_{\alpha}$ denote the image of $\alpha$ under a fixed isomorphism from $A$ to $\widehat{A}$. The trivial character $\chi_{0}$ of $A$ is defined by $\chi_{0}(a)=1$ for every $a\in A$. In particular, if $A=\mathbb{F}_{q}$ is the finite field with $q$ elements, then all additive characters of $\mathbb{F}_{q}$ are given by
\[
\chi_{\alpha}:\mathbb{F}_{q}\longrightarrow\mathbb{C}^{*},
\qquad
\chi_{\alpha}(x)=\zeta_{p}^{\Tr_{q/p}(\alpha x)}
\quad(\alpha\in\mathbb{F}_{q}),
\]
where $\Tr_{q/p}$ is the trace function from $\mathbb{F}_{q}$ to $\mathbb{F}_{p}$, and $\zeta_{k}$ is a primitive complex $k$-th root of unity. If $A=\mathbb{F}_{q}^{*}$, then all multiplicative characters of $\mathbb{F}_{q}$ are given by
\[
\{\chi^{i}:0\leq i\leq q-2\},
\]
where $\chi(\gamma)=\zeta_{q-1}$ and $\mathbb{F}_{q}^{*}=\langle\gamma\rangle$.

For every $a\in A^{*}$, it is easy to see that
\[
\sum_{\alpha\in A}\chi_{\alpha}(a)=0.
\]
The orthogonality relations for characters are
\begin{equation*}
\sum_{a\in A}
\chi_{\alpha_{1}}(a)\overline{\chi_{\alpha_{2}}(a)}
=
\begin{cases}
0, & \text{if }\alpha_{1}\neq\alpha_{2},\\[2pt]
|A|, & \text{if }\alpha_{1}=\alpha_{2}.
\end{cases}
\end{equation*}

\subsection{Discrete Fourier transform}

Let $\varphi$ be any complex-valued function on $A$. Its discrete Fourier transform is defined by
\[
\widehat{\varphi}(\alpha)
=
\sum_{a\in A}\varphi(a)\chi_{\alpha}(a).
\]
The inverse Fourier transform of $\varphi$ is
\[
\varphi(a)
=
\frac{1}{|A|}
\sum_{\alpha\in A}
\widehat{\varphi}(\alpha)\overline{\chi_{\alpha}(a)},
\qquad a\in A.
\]

\begin{lemma}[Parseval's identity \cite{carlet2004}]\label{lem:parseval}
The following identity holds:
\[
\sum_{a\in A}|\varphi(a)|^{2}
=
\frac{1}{|A|}
\sum_{\alpha\in A}|\widehat{\varphi}(\alpha)|^{2}.
\]
\end{lemma}

Let $f$ be a function from $A$ to a finite Abelian group $B$. Fix an isomorphism between $B$ and $\widehat{B}$, the group of all characters of $B$, and let $\psi_{\beta}$ denote the image of $\beta\in B$ under this isomorphism. For each $\beta\in B$, set $f_{\beta}:=\psi_{\beta}\circ f$. Then
\[
\widehat{f_{\beta}}(\alpha)
=
\sum_{a\in A}
(\psi_{\beta}\circ f)(a)\chi_{\alpha}(a).
\]
By Parseval's identity in Lemma~\ref{lem:parseval},
\[
\sum_{\alpha\in A}|\widehat{f_{\beta}}(\alpha)|^{2}=|A|^{2}.
\]
Since the root mean square does not exceed the maximum, we obtain
\[
\max_{\alpha\in A}|\widehat{f_{\beta}}(\alpha)|\geq\sqrt{|A|}.
\]

Further details on the Fourier transform can be found in \cite[Section 3.4]{carlet2004}.

\subsection{Perfect nonlinear functions}

Let $f$ be a function from an Abelian group $(A,+)$ to another Abelian group $(B,+)$, and define
\[
D_{a}f(x)=f(x+a)-f(x).
\]
The nonlinearity of $f$ is measured by
\[
P_{f}
=
\max_{a\in A^{*}}\max_{b\in B}
\Pr\bigl(D_{a}f(x)=b\bigr),
\]
where $\Pr(E)$ denotes the probability that event $E$ occurs. The smaller the value of $P_{f}$, the higher the nonlinearity of $f$. Functions with high nonlinearity have important applications in cryptography, sequence design, and coding theory.

\begin{definition}\label{def:perfect-nonlinear}
Let $f:A\to B$ be a function from an Abelian group $A$ to an Abelian group $B$. If
\[
P_{f}=\frac{1}{m},
\]
where $|B|=m$, then $f$ is called a perfect nonlinear function.
\end{definition}

For perfect nonlinear functions on Abelian groups, Carlet and Ding studied the Fourier transform and obtained the following result \cite{carlet2004}.

\begin{lemma}\label{lem:carlet-ding}
\cite[Theorem 16]{carlet2004}
Let $f:A\to B$ be a function from an Abelian group $A$ to an Abelian group $B$. Then $f$ is perfect nonlinear if and only if
\[
|\widehat{f_{\beta}}(\alpha)|=\sqrt{n}
\]
for every $\beta\in B^{*}$ and every $\alpha\in A$, where $|A|=n$.
\end{lemma}
Such functions are also called bent.
\section{A generic construction and a modified construction of codebooks}

In this section, we present two constructions of codebooks and continue to use the notation introduced in Section~2.

Let $f$ be a fixed function from a finite Abelian group $A$ to another finite Abelian group $B$, where $|A|=n$ and $|B|=m$. We consider the nontrivial case $n>1$ throughout. Let $\mathcal{E}_{n}$ be the set of standard basis vectors of the $n$-dimensional Hilbert space:
\[
\begin{aligned}
&(1,0,0,\ldots,0,0),\\
&(0,1,0,\ldots,0,0),\\
&\qquad\vdots\\
&(0,0,0,\ldots,0,1).
\end{aligned}
\]

\subsection{The generic construction}

For every $\chi_{\alpha}\in\widehat{A}$ and every $\psi_{\beta}\in\widehat{B}$, define the following codeword of length $n$:
\begin{equation*}
\bm{c}_{(\alpha,\beta)}
=
\frac{1}{\sqrt{n}}
\bigl(f_{\beta}(a)\chi_{\alpha}(a)\bigr)_{a\in A},
\end{equation*}
where $f_{\beta}=\psi_{\beta}\circ f$. This codeword is a unit-norm $1\times n$ complex vector.

We now construct a complex codebook $\mathcal{C}$ as follows:
\begin{equation}\label{eq:first-codebook}
\mathcal{C}
=
\{\bm{c}_{(\alpha,\beta)}:\alpha\in A,\ \beta\in B\}
\cup
\mathcal{E}_{n}.
\end{equation}
Write
\[
\mathcal{S}_{f}
=
\{\bm{c}_{(\alpha,\beta)}:\alpha\in A,\ \beta\in B\},
\]
and define the parametrization
\[
\Phi_{f}:A\times B\longrightarrow\mathcal{S}_{f},
\qquad
(\alpha,\beta)\longmapsto\bm{c}_{(\alpha,\beta)}.
\]
For a general function $f$, the map $\Phi_{f}$ need not be injective; that is, distinct parameter pairs may yield the same codeword. Every coordinate of each codeword in $\mathcal{S}_{f}$ has modulus $1/\sqrt{n}$, whereas every vector in $\mathcal{E}_{n}$ has a zero coordinate. Hence, for $n>1$,
\[
\mathcal{S}_{f}\cap\mathcal{E}_{n}=\varnothing.
\]
Therefore,
\[
N=|\mathcal{C}|=|\mathcal{S}_{f}|+n\leq mn+n,
\]
with equality if and only if $\Phi_{f}$ is injective.

\begin{theorem}\label{thm:first-construction}
Let $\mathcal{C}$ be the codebook in \eqref{eq:first-codebook}. Then $\mathcal{C}$ is an $(N,n)$ codebook, where
\[
N=|\mathcal{S}_{f}|+n\leq mn+n.
\]
Its maximum cross-correlation magnitude is
\begin{align*}
I_{\max}(\mathcal{C})
=
\max\Biggl\{
&\frac{1}{\sqrt{n}},\\
&\max\left\{
\frac{1}{n}
\left|
\widehat{f_{\beta_{1}-\beta_{2}}}
(\alpha_{1}-\alpha_{2})
\right|:
\begin{array}{l}
(\alpha_{i},\beta_{i})\in A\times B,\ i=1,2,\\
\bm{c}_{(\alpha_{1},\beta_{1})}
\neq
\bm{c}_{(\alpha_{2},\beta_{2})}
\end{array}
\right\}
\Biggr\}.
\end{align*}
If the inner set is empty, its maximum is understood to be $0$.
\end{theorem}

\begin{proof}
Let $\bm{c}_{1}$ and $\bm{c}_{2}$ be two distinct codewords of $\mathcal{C}$. We consider the following three cases.

(1) If $\bm{c}_{1},\bm{c}_{2}\in\mathcal{E}_{n}$, then
$
|\bm{c}_{1}\bm{c}_{2}^{H}|=0.
$

(2) If one codeword belongs to $\mathcal{S}_{f}$ and the other belongs to $\mathcal{E}_{n}$, then for some $a\in A$ and some $(\alpha,\beta)\in A\times B$,
\[
|\bm{c}_{1}\bm{c}_{2}^{H}|
=
\frac{1}{\sqrt{n}}
|f_{\beta}(a)\chi_{\alpha}(a)|
=
\frac{1}{\sqrt{n}}.
\]

(3) If $\bm{c}_{1},\bm{c}_{2}\in\mathcal{S}_{f}$, write
$
\bm{c}_{1}=\bm{c}_{(\alpha_{1},\beta_{1})}$, 
$
\bm{c}_{2}=\bm{c}_{(\alpha_{2},\beta_{2})}.
$
Since $\bm{c}_{1}\neq\bm{c}_{2}$, these two parameter representations correspond to two genuinely distinct codewords. Put
\[
\alpha=\alpha_{1}-\alpha_{2},
\qquad
\beta=\beta_{1}-\beta_{2}.
\]
Then
\[
\begin{aligned}
\bm{c}_{1}\bm{c}_{2}^{H}
&=
\frac{1}{n}
\sum_{a\in A}
(\psi_{\beta_{1}}\overline{\psi_{\beta_{2}}})\circ f(a)
(\chi_{\alpha_{1}}\overline{\chi_{\alpha_{2}}})(a)\\
&=
\frac{1}{n}
\sum_{a\in A}\psi_{\beta}\circ f(a)\chi_{\alpha}(a)\\
&=
\frac{1}{n}\widehat{f_{\beta}}(\alpha).
\end{aligned}
\]
Combining the three cases gives the stated formula. In particular, if $\Phi_{f}$ is injective, then $N=mn+n$ and distinct parameter pairs correspond to distinct codewords.
\end{proof}

By Theorem~\ref{thm:first-construction}, the maximum cross-correlation magnitude $I_{\max}(\mathcal{C})$ is determined by the Fourier transform of $f$. Thus, by choosing suitable groups $A$ and $B$ and a suitable function $f$, we can obtain codebooks with good parameters.

We next obtain optimal codebooks by choosing perfect nonlinear functions.

\begin{theorem}\label{thm:perfect-nonlinear}
Let $f$ be a perfect nonlinear function from $A$ to $B$. Then the following statements hold.
\begin{enumerate}
  \item[(1)] The codebook $\mathcal{C}$ in Theorem~\ref{thm:first-construction} has parameters $(mn+n,n)$ and
  \[
  I_{\max}(\mathcal{C})=\frac{1}{\sqrt{n}}.
  \]
  \item[(2)] The codebook $\mathcal{C}$ in Theorem~\ref{thm:first-construction} is optimal with respect to the Levenshtein bound if and only if $|A|=|B|$.
\end{enumerate}
\end{theorem}

\begin{proof}
We first show that the codewords in \eqref{eq:first-codebook} are distinct under the perfect nonlinearity condition. Suppose that
$
\bm{c}_{(\alpha_{1},\beta_{1})}
=
\bm{c}_{(\alpha_{2},\beta_{2})}.
$
Put $\alpha=\alpha_{1}-\alpha_{2}$ and $\beta=\beta_{1}-\beta_{2}$. Then, for every $a\in A$,
\[
\psi_{\beta}\circ f(a)\chi_{\alpha}(a)=1.
\]
If $\beta\neq0$, summing over $a\in A$ gives
\[
\widehat{f_{\beta}}(\alpha)=n,
\]
and hence
\[
|\widehat{f_{\beta}}(\alpha)|=n.
\]
On the other hand, Lemma~\ref{lem:carlet-ding} implies that a perfect nonlinear function satisfies
\[
|\widehat{f_{\beta}}(\alpha)|=\sqrt{n},
\]
which contradicts $n>1$. Therefore, $\beta=0$. The preceding identity then becomes
\[
\chi_{\alpha}(a)=1
\]
for every $a\in A$, and thus $\alpha=0$. Consequently,
\[
(\alpha_{1},\beta_{1})=(\alpha_{2},\beta_{2}),
\]
so $\Phi_{f}$ is injective. Since $\mathcal{S}_{f}\cap\mathcal{E}_{n}=\varnothing$, we obtain
\[
N=mn+n.
\]

We now determine the maximum cross-correlation magnitude. For two distinct codewords in $\mathcal{S}_{f}$, if $\beta_{1}-\beta_{2}\neq0$, then Lemma~\ref{lem:carlet-ding} gives a correlation magnitude of $1/\sqrt{n}$. If $\beta_{1}-\beta_{2}=0$, then necessarily $\alpha_{1}-\alpha_{2}\neq0$, and the character orthogonality relations give a correlation magnitude of $0$. Moreover, the correlation magnitude between any codeword in $\mathcal{S}_{f}$ and any standard basis vector in $\mathcal{E}_{n}$ is $1/\sqrt{n}$. Hence,
\[
I_{\max}(\mathcal{C})=\frac{1}{\sqrt{n}}.
\]

The codebook $\mathcal{C}$ is optimal with respect to the Levenshtein bound if and only if
\[
I_{\max}(\mathcal{C})
=
\frac{1}{\sqrt{K}}
=
\sqrt{
\frac{2N-K^{2}-K}{(N-K)(K+1)}
},
\qquad
N>K^{2},
\]
where $N=mn+n$ and $K=n$. Substituting these expressions yields $m=n$. Therefore, $\mathcal{C}$ is optimal with respect to the Levenshtein bound if and only if $|A|=|B|$.
\end{proof}

By Lemmas~\ref{lem:welch} and~\ref{lem:levenshtein}, if $f$ is a perfect nonlinear function from $A$ to $B$, then the codebook in Theorem~\ref{thm:perfect-nonlinear} is at least asymptotically optimal as $|B|\to\infty$.

In the remainder of this section, we use several classes of perfect nonlinear functions to obtain optimal or asymptotically optimal codebooks.

\begin{corollary}\label{cor:finite-field}
In Theorem~\ref{thm:perfect-nonlinear}, let
$
f:(\mathbb{F}_{q}^{l},+)\longrightarrow(\mathbb{F}_{q},+)
$
be a perfect nonlinear function, where $q$ is a power of a prime $p$ and $l$ is a positive integer. Then the resulting codebook $\mathcal{C}$ has the following properties.
\begin{enumerate}
  \item[(1)] If $l=1$, then $\mathcal{C}$ is an optimal codebook with parameters $(q^{2}+q,q)$ and
  \[
  I_{\max}(\mathcal{C})=\frac{1}{\sqrt{q}}.
  \]
  \item[(2)] If $l>1$, then $\mathcal{C}$ is an asymptotically optimal codebook with parameters $(q^{l+1}+q^{l},q^{l})$ and
  \[
  I_{\max}(\mathcal{C})=\frac{1}{\sqrt{q^{l}}}.
  \]
\end{enumerate}
\end{corollary}

\begin{proof}
If $l=1$, the result follows directly from Theorem~\ref{thm:perfect-nonlinear}.

Let $l>1$. By Theorem~\ref{thm:perfect-nonlinear}, $\mathcal{C}$ is a codebook with parameters
\[
(q^{l+1}+q^{l},q^{l})
\]
and
\[
I_{\max}(\mathcal{C})=\frac{1}{\sqrt{q^{l}}}.
\]
By Lemma~\ref{lem:welch}, the corresponding Welch bound is
\[
I_{W}
=
\sqrt{
\frac{q}{q^{l+1}+q^{l}-1}
}.
\]
As $q\to\infty$,
\[
\lim_{q\to\infty}
\frac{I_{\max}(\mathcal{C})}{I_{W}}
=
\lim_{q\to\infty}
\sqrt{1+q^{-1}-q^{-l-1}}
=
1.
\]
Therefore, the codebook is asymptotically optimal.
\end{proof}

\begin{remark}\label{rem:known-pn}
In Corollary~\ref{cor:finite-field}, the case $l=1$ gives a construction already obtained by Ding and Yin \cite{ding2007signal}. However, the case $l>1$ yields a new construction, and hence extends the results in \cite{ding2007signal}. For an odd prime $p$ and $q=p^{m}$, several known classes of perfect nonlinear functions from $\mathbb{F}_{q}$ to $\mathbb{F}_{q}$ are listed below:
\begin{enumerate}
  \item[$\bullet$] $f(x)=x^{2}$ (the classical case);
  \item[$\bullet$] $f(x)=x^{p^{k}+1}$, where $m/\gcd(m,k)$ is odd \cite{dembowski1968};
  \item[$\bullet$] $f(x)=x^{(3^{k}+1)/2}$, where $k$ is odd, $p=3$, and $\gcd(m,k)=1$ \cite{coulter1997};
  \item[$\bullet$] $f(x)=x^{10}-ux^{6}-u^{2}x^{2}$, where $p=3$ and $m$ is odd \cite{coulter1997,ding2006skew};
  \item[$\bullet$] $f(x)=x^{p^{s}+1}-u^{p^{k}-1}x^{p^{k}+p^{2k+s}}$, where $m=3k$, $\gcd(3,k)=1$, $k-s\equiv0\pmod{3}$, $s\neq k$, $u$ is a primitive element of $\mathbb{F}_{p^{3k}}$, and $m/\gcd(m,s)$ is odd \cite{zha2009binomials};
  \item[$\bullet$] $f(x)=x^{p^{2i}+p^{i}}+bx^{p^{k}+1}+cx^{p^{k}(p^{2i}+p^{i})}$, where $m=2k$, $0<i<m$, $m/\gcd(i,m)$ is odd, and $b,c\in\mathbb{F}_{q}$ satisfy
  \[
  cb^{p^{k}}\neq b,
  \qquad
  c^{p^{k}+1}=1,
  \qquad
  c^{(p^{k}+1)/\gcd(p^{k}+1,p^{i}+1)}\neq1
  \]
  \cite{zha2009polynomials};
  \item[$\bullet$] For even $m$, define $f$ on $\mathbb{F}_{p^{m}}$ by
  \[
  f(x)=bx^{p^{i}+1}+b^{p^{m/2}}x^{p^{m/2+i}+p^{m/2}}
  +cx^{p^{m/2+h}+p^{h}}
  +\sum_{j=0}^{m-1}r_{j}x^{p^{j+k}+p^{j}},
  \]
  where $c\notin\mathbb{F}_{p^{m/2}}$, $0\leq h\leq m-1$, and $b\neq\lambda^{p^{i}+1}$ for every $\lambda\in\mathbb{F}_{p^{m}}$. In addition, for every $j$, $r_{j}\in\mathbb{F}_{p^{k}}$, $0<i<m$, $\gcd(i,m)=t$, and $m/t$ is odd \cite{zha2009polynomials}.
\end{enumerate}

For $l>1$, known perfect nonlinear functions from $(\mathbb{F}_{q}^{l},+)$ to $(\mathbb{F}_{q},+)$ are listed below:
\begin{enumerate}
  \item[$\bullet$] The generalized Boolean function
  \[
  f(x_{1},\ldots,x_{l})
  =
  x_{1}x_{l/2+1}
  +x_{2}x_{l/2+2}
  +\cdots
  +x_{l/2}x_{l}
  \]
  is perfect nonlinear with $P_{f}=1/q$, where $l$ is even \cite{dillon1974}.
  \item[$\bullet$] Let $l$ be even and let $\pi$ be a bijection from $\mathbb{F}_{q}^{l/2}$ to $\mathbb{F}_{q}^{l/2}$. Denote its coordinate functions by $\pi_{1},\ldots,\pi_{l/2}$, and let $g$ be a function from $\mathbb{F}_{q}^{l/2}$ to $\mathbb{F}_{q}$. Then
  \[
  \begin{aligned}
  f(x_{1},\ldots,x_{l})
  ={}&
  x_{1}\pi_{1}(x_{l/2+1},\ldots,x_{l})
  +\cdots\\
  &+x_{l/2}\pi_{l/2}(x_{l/2+1},\ldots,x_{l})
  +g(x_{l/2+1},\ldots,x_{l})
  \end{aligned}
  \]
  is perfect nonlinear with $P_{f}=1/q$ \cite{carlet2004}.
  \item[$\bullet$] Let $l$ be even and identify $\mathbb{F}_{q}^{l/2}$ with $\mathbb{F}_{q^{l/2}}$. Let $g$ be any balanced function from $\mathbb{F}_{q^{l/2}}$ to $\mathbb{F}_{q}$. Then
  \[
  f(x,x')=g\bigl(x(x')^{q^{l/2}-2}\bigr),
  \qquad
  x,x'\in\mathbb{F}_{q}^{l/2},
  \]
  is perfect nonlinear with $P_{f}=1/q$ \cite{carlet2004}.
\end{enumerate}
Using the perfect nonlinear functions for $l>1$ described above and Corollary~\ref{cor:finite-field}, we obtain three classes of new asymptotically optimal codebooks.
\end{remark}

\begin{example}\label{ex:binary}
Let
$
f(x_{1},x_{2})=x_{1}x_{2}
$
be the perfect nonlinear function from $(\mathbb{F}_{2}^{2},+)$ to $(\mathbb{F}_{2},+)$, where $p=2$ and $l=2$. By Corollary~\ref{cor:finite-field},
\[
\begin{aligned}
\mathcal{C}
={}&
\bigl\{
[0.5,0.5,0.5,0.5],
[0.5,-0.5,0.5,-0.5],\\
&\quad
[0.5,0.5,-0.5,-0.5],
[0.5,-0.5,-0.5,0.5],\\
&\quad
[0.5,0.5,0.5,-0.5],
[0.5,-0.5,0.5,0.5],\\
&\quad
[0.5,0.5,-0.5,0.5],
[0.5,-0.5,-0.5,-0.5]
\bigr\}
\cup\mathcal{E}_{4}
\end{aligned}
\]
is a codebook with parameters $(12,4)$ and $I_{\max}=0.5$. This example has been verified by a Magma program.
\end{example}

We now choose the perfect nonlinear function
\[
f(x_{1},\ldots,x_{l})
=
x_{1}x_{l/2+1}
+x_{2}x_{l/2+2}
+\cdots
+x_{l/2}x_{l}.
\]
Some parameters of the codebooks constructed in Corollary~\ref{cor:finite-field} are listed in Table~\ref{tab:cor1}. As $q$ increases, the ratio $I_{\max}/I_{W}$ rapidly approaches $1$.

\begin{table}[htp]
  \centering
  \caption{Parameters of the codebooks in Corollary~\ref{cor:finite-field} based on
  $f(x_{1},\ldots,x_{l})
  =x_{1}x_{l/2+1}+x_{2}x_{l/2+2}+\cdots+x_{l/2}x_{l}$}
  \label{tab:cor1}
  \begin{tabular}{ccccccc}
    \toprule
    $q$ & $l$ & $(N,K)$ & $I_{\max}$ & $I_{W}$ & $1/\sqrt{q^{l}}$ & $I_{\max}/I_{W}$\\
    \midrule
    3 & 2 & $(36,9)$    & 0.3333 & 0.2928 & 0.3333 & 1.1386\\
    5 & 2 & $(150,25)$  & 0.2000 & 0.1832 & 0.2000 & 1.0917\\
    7 & 2 & $(392,49)$  & 0.1429 & 0.1338 & 0.1429 & 1.0680\\
    9 & 2 & $(810,81)$  & 0.1111 & 0.1055 & 0.1111 & 1.0531\\
    \bottomrule
  \end{tabular}
\end{table}

\begin{corollary}\label{cor:integer-group}
In Theorem~\ref{thm:perfect-nonlinear}, let
$
f:(\mathbb{Z}_{q}^{l},+)\longrightarrow(\mathbb{Z}_{q},+)
$
be a perfect nonlinear function, where $q$ is a power of a prime $p$ and $l$ is a positive integer. Then the codebook $\mathcal{C}$ has the following properties.
\begin{enumerate}
  \item[(1)] If $l=1$, then $\mathcal{C}$ is an optimal codebook with parameters $(q^{2}+q,q)$ and
  \[
  I_{\max}(\mathcal{C})=\frac{1}{\sqrt{q}}.
  \]
  \item[(2)] If $l>1$, then $\mathcal{C}$ is an asymptotically optimal codebook with parameters $(q^{l+1}+q^{l},q^{l})$ and
  \[
  I_{\max}(\mathcal{C})=\frac{1}{\sqrt{q^{l}}}.
  \]
\end{enumerate}
\end{corollary}

\begin{proof}
The proof is analogous to that of Corollary~\ref{cor:finite-field}.
\end{proof}

Several known classes of perfect nonlinear functions from $(\mathbb{Z}_{q}^{l},+)$ to $(\mathbb{Z}_{q},+)$ are recalled below. Carlet et al. showed that a class of generalized bent functions over $\mathbb{Z}_{q}$, where $q$ is a prime power and $l$ is even, gives rise to such functions \cite{carlet1999}. Carlet and Ding constructed two classes of perfect nonlinear functions from $(\mathbb{Z}_{q}^{2},+)$ to $(\mathbb{Z}_{q},+)$ \cite{carlet2004}. Feng proposed perfect nonlinear functions from $(\mathbb{Z}_{p^{2}}^{l},+)$ to $(\mathbb{Z}_{p^{2}},+)$ under the condition that either $l\geq p$, or $l<p$ and $l$ is composite or can be expressed as a sum of positive composite integers \cite{feng2009}. Carlet and Dubuc constructed several perfect nonlinear functions from $\mathbb{Z}_{p^{k}}^{2l}$ to $\mathbb{Z}_{p^{k}}$ with $l\geq k$ \cite{carlet1999}. Therefore, by Corollary~\ref{cor:integer-group}, each of these functions yields a new codebook.

We next construct asymptotically optimal codebooks from a more general class of perfect nonlinear functions than that in Corollary~\ref{cor:integer-group}.

\begin{corollary}\label{cor:general-integer}
In Theorem~\ref{thm:perfect-nonlinear}, let
$
f:(\mathbb{Z}_{v}^{l},+)\longrightarrow(\mathbb{Z}_{v}^{s},+)
$
be a perfect nonlinear function, where $v$, $l$, and $s$ are positive integers. Then the following statements hold.
\begin{enumerate}
  \item[(1)] If $l=s$, then $\mathcal{C}$ is an optimal codebook with respect to the Levenshtein bound, with parameters
  \[
  (v^{2l}+v^{l},v^{l})
  \]
  and
  \[
  I_{\max}(\mathcal{C})=\frac{1}{\sqrt{v^{l}}}.
  \]
  \item[(2)] If $l>s$, then $\mathcal{C}$ is an asymptotically optimal codebook with parameters
  \[
  (v^{l+s}+v^{l},v^{l})
  \]
  and
  \[
  I_{\max}(\mathcal{C})=\frac{1}{\sqrt{v^{l}}}.
  \]
\end{enumerate}
\end{corollary}

\begin{proof}
By Theorem~\ref{thm:perfect-nonlinear}, $\mathcal{C}$ is a codebook with parameters
\[
(v^{l+s}+v^{l},v^{l})
\]
and
\[
I_{\max}(\mathcal{C})=\frac{1}{\sqrt{v^{l}}}.
\]
If $l=s$, optimality follows directly from Theorem~\ref{thm:perfect-nonlinear}. By Lemma~\ref{lem:welch}, the corresponding Welch bound is
\[
I_{W}
=
\sqrt{
\frac{v^{s}}{v^{l+s}+v^{l}-1}
}.
\]
Hence,
\[
\frac{I_{\max}(\mathcal{C})}{I_{W}}
=
\sqrt{1+v^{-s}-v^{-l-s}}.
\]
Clearly, as $v\to\infty$ or $s\to\infty$,
\[
\frac{I_{\max}(\mathcal{C})}{I_{W}}=1.
\]
Thus, the codebook is asymptotically optimal with respect to the Welch bound.
\end{proof}

Except for the case $l=s=1$ with $v=q$, the asymptotically optimal codebooks in Corollary~\ref{cor:general-integer} have new parameters. We now list some known perfect nonlinear functions from $(\mathbb{Z}_{v}^{l},+)$ to $(\mathbb{Z}_{v}^{s},+)$. Zhang et al. proved that a perfect nonlinear function from $\mathbb{Z}_{4}^{l}$ to $\mathbb{Z}_{4}^{s}$ exists if and only if
$
s\leq\left\lfloor\frac{l}{2}\right\rfloor.
$
They also showed that if a perfect nonlinear function from $\mathbb{Z}_{q}^{l}$ to $\mathbb{Z}_{q}^{s}$ exists, then necessarily
$
s\leq\left\lfloor\frac{l}{2}\right\rfloor,
$
where $q=4t+2$ \cite{zhang2005}. In addition, Zhang et al. constructed perfect nonlinear functions from $\mathbb{Z}_{4}^{2l}$ to $\mathbb{Z}_{4}^{s}$ with $s\mid2l$ \cite{zhang2008}. They also constructed perfect nonlinear functions from $\mathbb{Z}_{p^{2}}^{l}$ to
$
\mathbb{Z}_{p^{2}}^{\left\lfloor r(p,l)/2\right\rfloor},
$
where $p$ is prime and
\[
r(p,l)
=
\max_{k\mid l}
k\left(
\left\lfloor\frac{l/k-2}{p}\right\rfloor+1
\right),
\]
with $l$ the extension degree of $GR(p^{2},l)$ \cite{zhang2008}.

We encourage the reader to explore new perfect nonlinear functions. Any new perfect nonlinear function can be used in Theorem~\ref{thm:perfect-nonlinear} to construct a new codebook with favorable parameters.

\subsection{The modified construction}

Let $\mathcal{F}$ be a family of functions from a finite Abelian group $A$ to a finite Abelian group $B$. For a fixed $\psi_{\beta}\in\widehat{B}$ with $\beta\neq0$, every $\chi_{\alpha}\in\widehat{A}$, and every $f\in\mathcal{F}$, define the following codewords of length $n$:
\[
\bm{c}_{(\alpha,f)}
=
\frac{1}{\sqrt{n}}
\bigl(f_{\beta}(a)\chi_{\alpha}(a)\bigr)_{a\in A},
\qquad
\bm{c}_{(\alpha,0)}
=
\frac{1}{\sqrt{n}}
\bigl(\chi_{\alpha}(a)\bigr)_{a\in A},
\]
where $f_{\beta}=\psi_{\beta}\circ f$. Each codeword is a unit-norm $1\times n$ complex vector.

We now construct a complex codebook $\mathcal{C}$ as follows:
\begin{equation}\label{eq:second-codebook}
\mathcal{C}
=
\{\bm{c}_{(\alpha,f)}:\alpha\in A,\ f\in\mathcal{F}\}
\cup
\{\bm{c}_{(\alpha,0)}:\alpha\in A\}
\cup
\mathcal{E}_{n}.
\end{equation}
Let $f_{0}$ be the zero map from $A$ to $B$, and set
\[
\bm{c}_{(\alpha,f_{0})}=\bm{c}_{(\alpha,0)}.
\]
Write
\[
\widetilde{\mathcal{F}}
=
\mathcal{F}\cup\{f_{0}\},
\qquad
\mathcal{T}_{\mathcal{F}}
=
\{\bm{c}_{(\alpha,f)}:
\alpha\in A,\ f\in\widetilde{\mathcal{F}}\}.
\]
For a general family $\mathcal{F}$, distinct pairs $(\alpha,f)$ may produce the same codeword, and $\mathcal{F}$ itself may contain the zero map. Every coordinate of each codeword in $\mathcal{T}_{\mathcal{F}}$ has modulus $1/\sqrt{n}$, while every vector in $\mathcal{E}_{n}$ has a zero coordinate. Therefore, for $n>1$,
\[
\mathcal{T}_{\mathcal{F}}\cap\mathcal{E}_{n}=\varnothing.
\]
Hence,
\[
N=|\mathcal{C}|=|\mathcal{T}_{\mathcal{F}}|+n
\leq n|\widetilde{\mathcal{F}}|+n
\leq n|\mathcal{F}|+2n.
\]
In particular,
\[
N=n|\mathcal{F}|+2n
\]
only when $f_{0}\notin\mathcal{F}$ and the map
\[
(\alpha,f)\longmapsto\bm{c}_{(\alpha,f)},
\qquad
A\times\widetilde{\mathcal{F}}\longrightarrow\mathcal{T}_{\mathcal{F}},
\]
is injective.

For convenience, set
\[
f_{\beta,i}=\psi_{\beta}\circ f_{i},
\qquad
g_{i,j}=f_{i}-f_{j},
\qquad
g_{\beta,i,j}=\psi_{\beta}\circ g_{i,j}.
\]

\begin{theorem}\label{thm:second-construction}
Let $\mathcal{C}$ be the codebook in \eqref{eq:second-codebook}. Then $\mathcal{C}$ is an $(N,n)$ codebook, where
\[
N=|\mathcal{T}_{\mathcal{F}}|+n\leq n|\mathcal{F}|+2n.
\]
Its maximum cross-correlation magnitude is
\begin{align*}
I_{\max}(\mathcal{C})
=
\max\Biggl\{
&\frac{1}{\sqrt{n}},\\
&\max\left\{
\frac{1}{n}
\left|
\widehat{g_{\beta,i,j}}
(\alpha_{i}-\alpha_{j})
\right|:
\begin{array}{l}
(\alpha_{i},f_{i}),(\alpha_{j},f_{j})
\in A\times\widetilde{\mathcal{F}},\\
\bm{c}_{(\alpha_{i},f_{i})}
\neq
\bm{c}_{(\alpha_{j},f_{j})}
\end{array}
\right\}
\Biggr\}.
\end{align*}
If the inner set is empty, its maximum is understood to be $0$.
\end{theorem}

\begin{proof}
Let $\bm{c}_{1}$ and $\bm{c}_{2}$ be two distinct codewords of $\mathcal{C}$. We consider the following three cases.

(1) If $\bm{c}_{1},\bm{c}_{2}\in\mathcal{E}_{n}$, then
$
|\bm{c}_{1}\bm{c}_{2}^{H}|=0.
$

(2) If one codeword belongs to $\mathcal{T}_{\mathcal{F}}$ and the other belongs to $\mathcal{E}_{n}$, then
\[
|\bm{c}_{1}\bm{c}_{2}^{H}|=\frac{1}{\sqrt{n}}.
\]

(3) If $\bm{c}_{1},\bm{c}_{2}\in\mathcal{T}_{\mathcal{F}}$, write
$
\bm{c}_{1}=\bm{c}_{(\alpha_{i},f_{i})}$,
$
\bm{c}_{2}=\bm{c}_{(\alpha_{j},f_{j})},
$
where $f_{i},f_{j}\in\widetilde{\mathcal{F}}$. Put $\alpha=\alpha_{i}-\alpha_{j}$. Then
\[
\begin{aligned}
\bm{c}_{1}\bm{c}_{2}^{H}
&=
\frac{1}{n}
\sum_{a\in A}
\bigl(\psi_{\beta}\circ f_{i}(a)\bigr)
\overline{\bigl(\psi_{\beta}\circ f_{j}(a)\bigr)}
(\chi_{\alpha_{i}}\overline{\chi_{\alpha_{j}}})(a)\\
&=
\frac{1}{n}
\sum_{a\in A}
\psi_{\beta}\circ(f_{i}(a)-f_{j}(a))
\chi_{\alpha}(a)\\
&=
\frac{1}{n}
\widehat{g_{\beta,i,j}}(\alpha).
\end{aligned}
\]
Combining the three cases gives the stated formula. It is important to note that the inner maximum is taken only over genuinely distinct codewords, so different parameter representations of the same codeword are not counted as cross-correlations.
\end{proof}

\begin{theorem}\label{thm:family-conditions}
Let $\mathcal{F}$ be a family of functions from a finite Abelian group $A$ to a finite Abelian group $B$ satisfying the following conditions:
\begin{enumerate}
  \item[(1)] every function in $\mathcal{F}$ is a perfect nonlinear function;
  \item[(2)] the difference of any two distinct functions in $\mathcal{F}$ is also a perfect nonlinear function.
\end{enumerate}
Then the following statements hold.
\begin{enumerate}
  \item[(1)] The codebook $\mathcal{C}$ in Theorem~\ref{thm:second-construction} has parameters
  \[
  (n|\mathcal{F}|+2n,n)
  \]
  and
  \[
  I_{\max}(\mathcal{C})=\frac{1}{\sqrt{n}}.
  \]
  \item[(2)] The family $\mathcal{F}$ satisfies
  \[
  |\mathcal{F}|\leq n-1.
  \]
  \item[(3)] The codebook $\mathcal{C}$ in Theorem~\ref{thm:second-construction} is optimal with respect to the Levenshtein bound if and only if
  \[
  |\mathcal{F}|=|A|-1=n-1.
  \]
  \item[(4)] As $|\mathcal{F}|\to\infty$, the codebook $\mathcal{C}$ in Theorem~\ref{thm:second-construction} is asymptotically optimal with respect to the Welch bound.
\end{enumerate}
\end{theorem}

\begin{proof}
We first show that the conditions in Theorem~\ref{thm:family-conditions} ensure that the codewords in \eqref{eq:second-codebook} are distinct. Let $f_{i},f_{j}\in\widetilde{\mathcal{F}}$ and suppose that
\[
\bm{c}_{(\alpha_{i},f_{i})}
=
\bm{c}_{(\alpha_{j},f_{j})}.
\]
Put
\[
\alpha=\alpha_{i}-\alpha_{j},
\qquad
g=f_{i}-f_{j}.
\]
It follows from the equality of the codewords that, for every $a\in A$,
\[
\psi_{\beta}\circ g(a)\chi_{\alpha}(a)=1.
\]
Hence,
\[
\left|\widehat{g_{\beta}}(\alpha)\right|=n.
\]

If $f_{i}\neq f_{j}$, then two cases arise. If $f_{i},f_{j}\in\mathcal{F}$, condition (2) implies that $g=f_{i}-f_{j}$ is a perfect nonlinear function. If exactly one of $f_{i}$ and $f_{j}$ is $f_{0}$, then $g=\pm f$ for some $f\in\mathcal{F}$. Since perfect nonlinearity is preserved under negation, $g$ is again perfect nonlinear. By Lemma~\ref{lem:carlet-ding},
\[
\left|\widehat{g_{\beta}}(\alpha)\right|=\sqrt{n},
\]
which contradicts $n>1$. Therefore, $f_{i}=f_{j}$. In this case, $g=f_{0}$, and the preceding identity becomes
\[
\chi_{\alpha}(a)=1
\]
for every $a\in A$. Thus, $\alpha=0$, so $\alpha_{i}=\alpha_{j}$. Consequently, the map
\[
A\times\widetilde{\mathcal{F}}
\longrightarrow
\mathcal{T}_{\mathcal{F}},
\qquad
(\alpha,f)\longmapsto\bm{c}_{(\alpha,f)},
\]
is injective.

Since $\beta\neq0$ in the second construction, $B$ is a nontrivial group, that is, $|B|=m>1$. For the zero map $f_{0}$, we have
\[
P_{f_{0}}=1\neq\frac{1}{m}.
\]
Hence, $f_{0}$ is not a perfect nonlinear function. Condition (1) therefore implies that $f_{0}\notin\mathcal{F}$. Consequently,
\[
|\mathcal{T}_{\mathcal{F}}|
=
n(|\mathcal{F}|+1),
\qquad
N=n|\mathcal{F}|+2n.
\]

We now compute the maximum cross-correlation magnitude. For two distinct codewords in $\mathcal{T}_{\mathcal{F}}$, if $f_{i}\neq f_{j}$, then the above discussion shows that $g=f_{i}-f_{j}$ is a perfect nonlinear function. By Lemma~\ref{lem:carlet-ding}, the correlation magnitude is $1/\sqrt{n}$. If $f_{i}=f_{j}$, then necessarily $\alpha_{i}\neq\alpha_{j}$, and the character orthogonality relations imply that the correlation magnitude is $0$. In addition, the correlation magnitude between any codeword in $\mathcal{T}_{\mathcal{F}}$ and any standard basis vector in $\mathcal{E}_{n}$ is $1/\sqrt{n}$. Therefore,
\[
I_{\max}(\mathcal{C})=\frac{1}{\sqrt{n}}.
\]
This proves (1).

We next prove (2). Let $t=|\mathcal{F}|$ and suppose, for contradiction, that $t>n-1$. Since $t$ is an integer, $t\geq n$. Then
\[
N=n(t+2)>n^{2}=K^{2},
\]
so the Levenshtein bound for complex codebooks applies. By Lemma~\ref{lem:levenshtein},
\[
I_{\max}^{2}(\mathcal{C})
\geq
\frac{2N-K^{2}-K}{(N-K)(K+1)}.
\]
Substituting
\[
N=n(t+2),
\qquad
K=n,
\]
we obtain
\[
I_{L}^{2}
=
\frac{2t+3-n}{(t+1)(n+1)}.
\]
On the other hand, by (1),
\[
I_{\max}^{2}(\mathcal{C})=\frac{1}{n}.
\]
A direct calculation gives
\[
I_{L}^{2}-\frac{1}{n}
=
\frac{(n-1)(t-n+1)}{n(t+1)(n+1)}.
\]
Since $n>1$ and $t>n-1$, this expression is strictly positive. Hence,
\[
I_{L}>I_{\max}(\mathcal{C}),
\]
which contradicts the Levenshtein bound. Therefore,
\[
|\mathcal{F}|\leq n-1.
\]
This proves (2).

We now prove (3). If $|\mathcal{F}|=n-1$, then
\[
N=n(n+1),
\qquad
K=n,
\]
and $N>K^{2}$. Substituting these values into the Levenshtein bound in \eqref{eq:levenshtein-complex}, we obtain
\[
I_{L}
=
\sqrt{
\frac{2N-K^{2}-K}{(N-K)(K+1)}
}
=
\frac{1}{\sqrt{n}}
=
I_{\max}(\mathcal{C}).
\]
Thus, $\mathcal{C}$ attains the Levenshtein bound.

Conversely, if $\mathcal{C}$ attains the Levenshtein bound in \eqref{eq:levenshtein-complex}, then
\[
\frac{1}{\sqrt{n}}
=
\sqrt{
\frac{2N-K^{2}-K}{(N-K)(K+1)}
},
\qquad
N>K^{2}.
\]
Substituting
\[
N=n|\mathcal{F}|+2n,
\qquad
K=n,
\]
yields
\[
|\mathcal{F}|=n-1.
\]
This proves (3).

Finally, we prove (4). The Welch bound corresponding to the codebook $\mathcal{C}$ is
\[
I_{W}
=
\sqrt{
\frac{|\mathcal{F}|+1}
{n|\mathcal{F}|+2n-1}
}.
\]
Therefore,
\[
\frac{I_{\max}(\mathcal{C})}{I_{W}}
=
\sqrt{
1+
\frac{n-1}{n(|\mathcal{F}|+1)}
}.
\]
By (2), $|\mathcal{F}|\leq n-1$, so when $|\mathcal{F}|\to\infty$, $n$ must also tend to infinity. Moreover,
\[
0
\leq
\frac{n-1}{n(|\mathcal{F}|+1)}
<
\frac{1}{|\mathcal{F}|+1}
\longrightarrow0.
\]
Hence,
\[
\lim_{|\mathcal{F}|\to\infty}
\frac{I_{\max}(\mathcal{C})}{I_{W}}
=1.
\]
Thus, as $|\mathcal{F}|\to\infty$, the codebook $\mathcal{C}$ is asymptotically optimal with respect to the Welch bound.
\end{proof}

Perfect nonlinear function families satisfying the conditions of Theorem~\ref{thm:family-conditions} do exist. For example, let
\[
\mathcal{F}
=
\{af:a\in\mathbb{F}_{q}^{*}\},
\]
where $f$ is a perfect nonlinear function from $\mathbb{F}_{q^l}$ to $\mathbb{F}_{q}$. It is easy to verify that $\mathcal{F}$ satisfies the conditions of Theorem~\ref{thm:family-conditions}. This yields a codebook with parameters $(q^{l+1}+q^l,q^l)$. In fact, Theorem~\ref{thm:family-conditions} generalizes the codebook construction of Zhou, Ding and Li in \cite{zhou2014}.

\begin{theorem}\label{thm:new-family}
Let $K=\mathbb{F}_{q^{m}}$,
$
A=(K^{r}\times K^{r},+)$,
$
B=(\mathbb{F}_{q},+),
$
where $m$ and $r$ are positive integers. For each $1\leq i\leq r$, let
$
\sigma_{i}:K^{*}\longrightarrow K^{*}
$
be a permutation. For every $a\in K^{*}$, define
\[
f_{a}(\bm{x},\bm{y})
=
\Tr_{q^{m}/q}
\left(
\sum_{i=1}^{r}\sigma_{i}(a)x_{i}y_{i}
\right),
\]
where
\[
\bm{x}=(x_{1},\ldots,x_{r}),
\qquad
\bm{y}=(y_{1},\ldots,y_{r})\in K^{r}.
\]
Then every function in
\[
\mathcal{F}=\{f_{a}:a\in K^{*}\}
\]
is perfect nonlinear, and the difference of any two distinct functions in $\mathcal{F}$ is also perfect nonlinear. Moreover,
\[
|\mathcal{F}|=q^{m}-1.
\]
\end{theorem}

\begin{proof}
Fix $a\in K^{*}$. Take any
\[
(\bm{u},\bm{v})\in A\setminus\{\bm{0}\}
\]
and any $c\in\mathbb{F}_{q}$. By definition,
\[
D_{(\bm{u},\bm{v})}
f_{a}(\bm{x},\bm{y})
=
\Tr_{q^{m}/q}
\left(
\sum_{i=1}^{r}
\sigma_{i}(a)
(v_{i}x_{i}+u_{i}y_{i}+u_{i}v_{i})
\right).
\]
Hence, the equation
\[
D_{(\bm{u},\bm{v})}
f_{a}(\bm{x},\bm{y})=c
\]
is equivalent to
\[
L(\bm{x},\bm{y})=d,
\]
where
\[
L(\bm{x},\bm{y})
=
\Tr_{q^{m}/q}
\left(
\sum_{i=1}^{r}
\sigma_{i}(a)(v_{i}x_{i}+u_{i}y_{i})
\right)
\]
and
\[
d
=
c-
\Tr_{q^{m}/q}
\left(
\sum_{i=1}^{r}
\sigma_{i}(a)u_{i}v_{i}
\right)
\in\mathbb{F}_{q}.
\]
Clearly, $L$ is an $\mathbb{F}_{q}$-linear map from $K^{2r}$ to $\mathbb{F}_{q}$. Since $(\bm{u},\bm{v})\neq\bm{0}$, there exists at least one index $i$ such that $u_{i}\neq0$ or $v_{i}\neq0$. Because $\sigma_{i}(a)\neq0$, the nondegeneracy of the trace pairing implies that $L$ is not the zero map, and hence $L$ is surjective. Therefore,
\[
\dim_{\mathbb{F}_{q}}\ker L=2mr-1.
\]
Thus, for every $d\in\mathbb{F}_{q}$, the equation $L(\bm{x},\bm{y})=d$ has exactly
\[
q^{2mr-1}
=
\frac{|A|}{|B|}
\]
solutions. Consequently, $f_{a}$ is perfect nonlinear.

Now take any $a,b\in K^{*}$ with $a\neq b$. Then
\[
(f_{a}-f_{b})(\bm{x},\bm{y})
=
\Tr_{q^{m}/q}
\left(
\sum_{i=1}^{r}
(\sigma_{i}(a)-\sigma_{i}(b))x_{i}y_{i}
\right).
\]
Since every $\sigma_{i}$ is a permutation,
\[
\sigma_{i}(a)-\sigma_{i}(b)\neq0,
\qquad
1\leq i\leq r.
\]
Repeating the preceding argument shows that $f_{a}-f_{b}$ is also perfect nonlinear.

Finally, if $a\neq b$ and $f_{a}=f_{b}$, then $f_{a}-f_{b}$ is the zero map, contradicting the fact that $f_{a}-f_{b}$ is perfect nonlinear. Therefore, the functions $f_{a}$ are pairwise distinct, and hence
\[
|\mathcal{F}|=|K^{*}|=q^{m}-1.
\]
\end{proof}

\begin{corollary}\label{cor:new-family}
In Theorem~\ref{thm:new-family}, let $r=2$,
\[
\sigma_{1}(a)=a,
\qquad
\sigma_{2}(a)=a^{-1}.
\]
Then
\[
\mathcal{F}
=
\left\{
f_{a}(x_{1},x_{2},y_{1},y_{2})
=
\Tr_{q^{m}/q}
\left(
ax_{1}y_{1}+a^{-1}x_{2}y_{2}
\right):
a\in\mathbb{F}_{q^{m}}^{*}
\right\}
\]
satisfies the conditions of Theorem~\ref{thm:family-conditions}. The codebook $\mathcal{C}$ obtained from the second construction is an asymptotically optimal codebook with parameters
\[
(q^{5m}+q^{4m},q^{4m})
\]
and
\[
I_{\max}(\mathcal{C})=\frac{1}{q^{2m}}.
\]
\end{corollary}

\begin{proof}
The maps
\[
a\longmapsto a,
\qquad
a\longmapsto a^{-1}
\]
are permutations of $\mathbb{F}_{q^{m}}^{*}$. By Theorem~\ref{thm:new-family}, every function in $\mathcal{F}$ is perfect nonlinear, the difference of any two distinct functions in $\mathcal{F}$ is perfect nonlinear, and
\[
|\mathcal{F}|=q^{m}-1.
\]
Here,
\[
n=|A|=q^{4m}.
\]
By Theorem~\ref{thm:family-conditions}, $\mathcal{C}$ is a codebook with parameters
\[
\begin{aligned}
(N,K)
&=
(n|\mathcal{F}|+2n,n)\\
&=
\bigl(q^{4m}(q^{m}+1),q^{4m}\bigr)\\
&=
(q^{5m}+q^{4m},q^{4m}),
\end{aligned}
\]
and
\[
I_{\max}(\mathcal{C})
=
\frac{1}{\sqrt{n}}
=
\frac{1}{q^{2m}}.
\]

By Lemma~\ref{lem:welch}, the corresponding Welch bound is
\[
I_{W}
=
\sqrt{
\frac{q^{m}}{q^{5m}+q^{4m}-1}
}.
\]
Therefore,
\[
\frac{I_{\max}(\mathcal{C})}{I_{W}}
=
\sqrt{
1+q^{-m}-q^{-5m}
}.
\]
As $q\to\infty$ or $m\to\infty$,
\[
\frac{I_{\max}(\mathcal{C})}{I_{W}}\longrightarrow1.
\]
Hence, the codebook is asymptotically optimal with respect to the Welch bound.
\end{proof}

The reader is encouraged to seek new families of perfect nonlinear functions satisfying the conditions of Theorem~\ref{thm:family-conditions}, thereby obtaining new optimal or asymptotically optimal codebooks.

\section{Conclusion}

This paper has presented two general constructions of complex codebooks from functions on finite Abelian groups and has computed their maximum cross-correlation magnitudes by means of the discrete Fourier transform. By choosing perfect nonlinear functions on finite Abelian groups, we have obtained conditions under which the constructed codebooks are optimal or nearly optimal. In particular, we have constructed several classes of optimal or asymptotically optimal codebooks with new parameters and generalized the results of Ding and Yin in \cite{ding2007signal} and Zhou, Ding and Li in \cite{zhou2014}.

\section*{Acknowledgments}
The authors sincerely thank the organizers and participants of SETA 2026 for valuable comments and discussions. This paper was accepted and presented at SETA 2026.
\thanks{This research was supported in part by the National Natural Science Foundation of China under Grant 12271059 and in part by the Natural Science Basic Research Program of Shaanxi (No. 2026JC-YXQN-030).}



\medskip
Received xxxx 20xx; revised xxxx 20xx; early access xxxx 20xx.
\medskip

\end{document}